\documentclass[11pt]{article}
\usepackage{epsf}
\usepackage{latexsym}
\font\mybb=msbm10 at 12pt
\font\mybbb=msbm10 at 9pt

\newcommand{\bZ}{\hbox{\mybb{Z}}}
\newcommand{\bN}{\hbox{\mybb{N}}}

\newcommand{\bR}{\hbox{\mybb{R}}}
\newcommand{\bC}{\hbox{\mybb{C}}}
\newcommand{\bc}{\hbox{\mybbb{C}}}
\newcommand{\bH}{\hbox{\mybb{H}}}
\newcommand{\bq}{\begin{equation}} 
\newcommand{\eq}{\end{equation}} 
\newcommand{\bqali}{\begin{eqnarray}}
\newcommand{\eqali}{\end{eqnarray}} 
\newcommand{\no}{\noindent}
\newcommand{\g}[1]{\mathsf{#1}}
\newtheorem{theorem}{Theorem}

\title{Homogeneous HKT and QKT manifolds}

\author{A. Opfermann\footnote{e-mail:ao200@damtp.cam.ac.uk}
\ and G. Papadopoulos\footnote{e-mail:gp204@damtp.cam.ac.uk}\\
\normalsize
\em Department of Applied Mathematics and Theoretical Physics, \\
\normalsize
\em University of Cambridge, \\
\normalsize
\em Silver Street, \\
\normalsize
\em Cambridge, CB3 9EW, UK}

\begin{document}

\maketitle

\begin{abstract}

We present the construction of a large class of homogeneous  KT, HKT and QKT
manifolds, $G/K$, using an invariant metric on 
$G$ and the canonical connection.
For this a  decomposition of the Lie algebra of $G$ is employed,
which is most
easily described  in terms of colourings of Dynkin diagrams of simple
Lie algebras. 
KT structures on homogeneous spaces 
are associated with different 
colourings of  Dynkin diagrams. The colourings  which 
give rise to HKT structures are found using extended Dynkin diagrams. 
We  also construct homogeneous QKT manifolds  from homogeneous HKT
manifolds and show that their twistor spaces  admit a KT
structure. Many examples of  homogeneous KT, HKT and QKT spaces are given.

\bigskip

\noindent DAMTP-1998-93

\end{abstract}

\vfill


\section{Introduction}

A hermitian manifold $M$ with complex structure 
$I$ and metric $B$ admits a K\"ahler 
structure with torsion (KT) provided that
\bq
\nabla I=0 \ ,
\label{con}
\eq
 where $\nabla$ is the connection
\bq
\nabla={\buildrel  o \over\nabla}+ H\ ,
\label{cona}
\eq
 ${\buildrel  o \over\nabla}$ is the Levi-Civita connection of $B$
and $H$ is  the torsion,  which is a three-form on $M$.   The condition
(\ref{con}) implies that the holonomy of the 
connection $\nabla$ is a subgroup
of $U(d)$, $\mathrm {dim}(M)=2\, d$. A KT 
structure is a generalization of a 
K\"ahler structure (K) and reduces to the latter 
whenever the torsion $H$ vanishes.

A hyper-complex tri-hermitian manifold $M$ with complex 
structures $\{I_r; r=1,2,3\}$
and metric $B$ admits a hyper-K\"ahler structure 
with torsion (HKT) provided that
\bq
\nabla I_r=0\ .
\label {aonea}
\eq
The holonomy of the connection  $\nabla$ 
is a subgroup of $Sp(d)$, ${\rm dim} (M)=4\, d$. 

In the same way an almost quaternionic tri-hermitian  
manifold $M$ with tangent bundle
endomorphisms $\{J_r; r=1,2,3\}$ and metric $B$ admits  a  quaternionic
K\"ahler 
structure with torsion (QKT) \cite{qkt} provided that
\bq
\nabla J_r= -2 \epsilon_{r}{}^{st}A_s J_t 
\label {aoneb}
\eq
and
\bq
N_D (J_r)=0 \ ,
\label {aonebb}
\eq
 where  $A$ is a $Sp(1)$-connection and $N_D (J_r)$ 
is the Nijenhuis type tensor
with respect to the covariant derivative $D$ of $A$. 
The holonomy of  the connection $\nabla$ is a 
subgroup of $Sp(d)\cdot Sp(1)$, ${\rm dim} (M)=4\, d$.
We remark that the definitions of QKT and of quaternionic
manifolds both utilize the existence
of the endomorphisms $\{J_r; r=1,2,3\}$. However, 
in the former case the relevant connection
is metric with torsion whereas in the
latter case the connection is torsion free.
A consequence of the conditions (\ref{con}), 
(\ref{aonea}), (\ref{aoneb}) and
(\ref{aonebb}) is that the torsion $H$ is
a (1,2)-
and (2,1)-form  with respect to  the endomorphisms $I$, $I_r$ and  $J_r$,
respectively. 
If the torsion $H$ is 
closed,  we say that the KT, HKT and QKT
structures are strong, otherwise  we say that they are weak \cite{hkt}. 
In what follows,
the KT, HKT and QKT structures which we shall consider are of the weak
type except  those which are on group manifolds.

The properties
of KT, HKT and QKT structures  closely resemble those
of  K, HK and QK ones, respectively. 
In particular, HKT \cite{hkt} and  QKT \cite{qkt} geometries
admit twistor constructions  with twistor spaces which
have similar properties to those of HK \cite{hk} and QK 
\cite{sala, ishi} manifolds.
Many examples of manifolds admitting  KT and HKT structures are known.
In particular, it was  shown in \cite{spin} that all
$2k$- and some $4k$-dimensional compact Lie groups
admit strong KT and HKT structures. The simplest such manifold with
an HKT structure is the Hopf surface $S^1\times S^3$.
In physics, KT, HKT and QKT manifolds arise as target
spaces of two-dimensional supersymmetric sigma models with Wess-Zumino term
\cite{sigma1, sigma2}, for which  
the  torsion is proportional to the Wess-Zumino term.
Another application of these geometries is in the context of
black-holes, where 
the moduli spaces of a class of black-hole supergravity solutions
are HKT manifolds.
Homogeneous K   \cite{bor, mat, roemer} and QK 
\cite{wolf} manifolds
have been investigated, and they have
found many applications in physics in the context
of sigma models and in supergravity theories \cite{wit}.
  
The present paper is dedicated to the investigation of 
KT, HKT and QKT structures
on homogeneous spaces $G/K$ which generalizes the construction of 
KT and HKT structures on group
manifolds.  Specifically, we  
show that the complex homogeneous spaces of \cite{wang} and 
hyper-complex homogeneous
spaces of \cite{joy}  admit   KT and  HKT  structures, respectively. 
In addition, we find homogeneous QKT spaces associated to a class of HKT ones.
In order to do this
we use an invariant metric on $G$ and the canonical connection
on $G/K$.  Our method to  construct homogeneous KT, HKT and QKT
spaces is based in part on the colouring  of  Dynkin
diagrams of simple Lie algebras. One of the advantages of this approach
is that it gives a simple description of a decomposition of the
Lie algebra of $G$ which is employed in the definition of the homogeneous
KT, HKT and QKT structures. It also allows us to compile lists
of such spaces. This includes the example of $S^1\times S^3$, which
admits an HKT structure, as well as a QK  structure.

This paper is organized as follows:  In section two,
we set up our notation. 
In section three,
we show that all  
homogeneous  complex  manifolds admit KT structures and compile
a list of such manifolds using Dynkin diagrams. In
section four, we give the construction of homogeneous HKT manifolds. 
In section five, we illustrate this construction 
employing Dynkin diagrams and present many examples of homogeneous HKT
manifolds. In section six,
we investigate homogeneous QKT manifolds. In section seven, we
show that the twistor spaces of QKT manifolds admit a KT structure, 
and in section
eight we give our conclusions.


\section{Lie groups  and homogeneous spaces}

Let $G$ be a  semi-simple, compact Lie group 
with  compact Lie algebra $\g{g}$  and $\g{g^c}=\g{g}\otimes 
\bc$.\footnote{In what follows, 
$\g{l^c}$ will denote the complexification, 
$\g{l^c}=\g{l}\otimes \bc$, of the
vector space $\g{l}$.} 
Let $\g{h}$ be a  Cartan sub-algebra of $\g{g}$,
$H$ be the associated maximal Abelian subgroup of $G$ and  $\Delta$
be the root system of $\g{g^c}$ with respect to $\g{h^c}$. Then 
(see for example \cite[p. 165]{hel}) 
\bq
\g{g^c}=\g{h^c} \ \oplus^{{\!\!\!\!\!}^{}}_
{{\!\!\!\!\!\!\!\!}_{_{\alpha \in \Delta}}} \, \g{g^c}_{(\alpha)}\ , \ 
\label{rootde}
\eq
   where  the root subspaces of $\g{g^c}$ are
\bq
\g{g}_{(\alpha)}^c=\{g_{\alpha} \in \g{g^c}:
[h,g_{\alpha}]=\alpha(h)g_{\alpha} \ , \forall \ h \in \g{h^c}\} \ .
\label{rootsub}
\eq
 We denote by $\Delta^+$ ($\Delta^-$)
the space of positive (negative) roots of $\g{g^c}$.
The choice of $\Delta^{+}$  
is in one-to-one correspondence with the choice
of a Weyl chamber  in $\g{h^c}$ \cite[p. 458]{hel}. 
Any root of $\g{g^c}$ can be written as the sum 
of simple roots, $\Delta^s:= \{\alpha_i:i=1,\ldots,
l=\mathrm{rank}(\g{g})\}$, with positive  integer coefficients
\cite[p. 177]{hel}.
Furthermore, the highest root $\psi$ in $\Delta^+$  
is unique for every simple
Lie algebra. 

For each linear function $\alpha$ on $\g{h^c}$
there exists a unique element $\tilde{h}_{\alpha}$ of $\g{h^c}$ 
such that the Killing metric
$B$  induces an inner product on the root space 
\cite[p. 500]{corn}
\bq
\alpha \cdot \beta := \alpha({\tilde{h}_{\beta}}) = 
B(\tilde{h}_{\alpha},\tilde{h}_{\beta}) \ . 
\eq
 Then the commutation relations of $\g{g^c}$ in the 
Chevalley basis \cite{che} are 
\bqali
[h_{\alpha}, e_{\beta}] &=& 2 \frac{\alpha \cdot \beta}
{\alpha \cdot \alpha} e_{\beta} \ ,\label{algebra1}\\
{[e_{\alpha}, e_{\beta}]} &=& N_{\alpha ,\beta}\, e_{\alpha +\beta} +
\delta_{\alpha ,-\beta}\, h_{\alpha}  \ ,
\label{algebra2}
\eqali
 where  $\{e_{\alpha}\}$ are the 
step operators, $\{h_\alpha\}$ are the generators
of the Cartan sub-algebra and
  the structure
constants $N_{\alpha,\beta}$ are integers. In particular,
\bq
N_{\alpha,\beta}= \pm (p+1)  \ ,
\label{stru}
\eq
 where  $-p \leq  n \leq q$ 
for the maximal $\alpha$-string, $\beta + n
\alpha$, containing $\beta$.  We remark that  \cite[p. 171]{hel}
\bq
N_{\alpha, \beta} = N_{\beta, -\alpha - \beta} = N_{-\alpha - \beta, \alpha} 
= - N_{\beta, \alpha}= - N_{-\alpha, -\beta} \ .
\label{ident}
\eq
 The Killing metric in the Chevalley basis 
$\{e_{\alpha}, h_{\alpha_i} \}$ is 
\bq
B = \left(\begin{array}{cc}
\frac{2}{\alpha \cdot \alpha} \ \delta_{\alpha, -\beta} & 0\\
 0 &  \frac{2}{\alpha_j \cdot \alpha_j} A_{ij} 
\end{array}\right) \ ,
\label{killcom}
\eq
 where $\alpha_i\in \Delta^s$ and 
\bq
A_{ij}= 2 \frac{\alpha_i \cdot \alpha_j}{\alpha_i \cdot
\alpha_i}
\eq
 is the Cartan matrix of $\g{g^c}$. It is 
well known that the Cartan matrix can be  
represented  pictorially  by  a Dynkin 
diagram \cite[p. 462]{hel}.  In the next sections 
we shall make extensive use of Dynkin 
diagrams in our 
construction of   homogeneous KT, HKT and QKT  manifolds.

A semi-simple complex Lie algebra 
$\g{g^c}$ admits a unique,    up to an  isomorphism,
compact real form $\g{g}$ \cite[p. 181 and 426]{hel}
\bq
E_{\alpha}^+= i(e_{\alpha} + e_{-\alpha}) \, , \ 
E_{\alpha}^-= (e_{\alpha} - e_{-\alpha}) \, , \
H_{\alpha}  = -i h_{\alpha} 
\label{compact}
\eq
  for  $\alpha \in \Delta^+$. Then the  commutation relations are  
\bqali
[H_{\alpha}, E_{\beta}^{\pm}] &=& \pm 2 \frac{\alpha \cdot \beta}{\alpha
\cdot \alpha} E_{\alpha}^{\mp} \ ,\label{newcom1}\\
{[E_{\alpha}^{\pm}, E_{\beta}^{\pm}]} &=& \mp N_{\alpha,\beta}
E_{\alpha + \beta}^- - N_{\alpha,-\beta} E_{\alpha - \beta}^-  \ ,
\label{newcom2}\\
{[E_{\alpha}^{\pm}, E_{\beta}^{\mp}]} &=& + N_{\alpha,\beta}
E_{\alpha + \beta}^+ \mp N_{\alpha,-\beta} E_{\alpha - \beta}^+ \pm  2
\delta_{\alpha,\beta} H_{\alpha} \ , 
\label{newcom3}
\eqali
 where the structure constants $N_{\alpha, \beta}$ are given in
(\ref{stru}). In the above expression of the commutators we have assumed 
that, if $\alpha-\beta\in \Delta$, then
$\alpha-\beta \in \Delta^+$. But if $\alpha-\beta \in \Delta^-$,
then $\beta-\alpha \in \Delta^+$
and the commutators can be easily re-expressed.

Let $\g{g}$ be a reductive Lie algebra, 
i.e. $\g{g}=\g{\tilde{u}}\,\oplus\, \g{\tilde{h}_0}$ is the direct sum
of the semi-simple Lie algebra $\g{\tilde{u}}$ and the Abelian Lie algebra 
$\g{\tilde{h}_0}$ 
\cite[p. 326]{nom}.  
Every compact Lie group
has a reductive Lie algebra.
An invariant metric on the compact Lie group $G$  restricted
on the semi-simple sub-algebra $\g{\tilde{u}}$ is proportional 
to the Killing metric on
$\g{\tilde{u}}$ and the Abelian Lie algebra $\g{\tilde{h}_0}$ is 
the orthogonal complement of
$\g{\tilde{u}}$ in
$\g{g}$.  

There is a basis  $\{E_{\alpha}^+, E_{\alpha}^-, H_{\alpha_i}, U_a\}$ 
in $\g{g}$ such that
any invariant metric $B$ on $\g{g}$ is
\bq
B = \left(\begin{array}{cccc}
 \frac{4}{\alpha \cdot \alpha} \delta_{\alpha \beta} & 0 & 0 & 0\\
 0 & \frac{4}{\alpha \cdot \alpha} \delta_{\alpha \beta} & 0  & 0\\
 0 & 0 & \frac{2}{\alpha_j \cdot \alpha_j} A_{ij} & 0\\
 0 & 0 & 0 & c_a \delta_{a b}
\end{array}\right) \ ,  
\label{killcom1}
\eq
 where  $\{c_a\}$ are  positive real constants and
$\{U_a\}$ is a basis in the ideal
$\g{h_0}$.

Let $M=G/K$  be a homogeneous manifold with $G$  a
compact Lie group  and $K$ a   closed and
connected subgroup of $G$. A  homogeneous space 
is reductive \cite[p.
190]{nom}  if there exists a subspace $\g{m}$ of $\g{g}$ such that 
\bq
\g{g} = \g{m} + \g{k}\, , \  \ [\g{k},\g{m}]
\subset \g{m} \ .
\label{comm2}
\eq
 If $\g{g}$ is equipped with an invariant metric $B$, then we define
\bq
\g{m} = \g{k}^{\perp} 
\label{perp}
\eq
 and the decomposition of $\g{g}$ is orthogonal, 
$\g{g} = \g{m} \oplus \g{k}$.
The invariant metric on $\g{g}$ induces an $\mathrm{ad}(G)$-invariant  
metric on $\g{m}$, the 
canonical normal metric \cite{zill} or normal homogeneous metric
\cite{wazi}, which in turn induces  
an invariant metric on $G/K$,  the standard homogeneous metric 
\cite{wazi}. We remark, that a
homogeneous space is  symmetric \cite[p. 226]{nom} if 
\bq
[\g{m},\g{m}] \subset \g{k} \ . 
\label{comm3}
\eq
Let $\g{k}$ be a sub-algebra of $\g{g}$ associated with a reductive
homogeneous space. The set of roots $\Delta$ of $\g{g}$ decomposes
as $\Delta=\Delta_{\g{m}} \cup \Delta_{\g{k}}$ with $\Delta_{\g{m}} \cap
\Delta_{\g{k}}=\emptyset$, where $\Delta_{\g{k}}$ is the root space of
${\g{k}}$ and  $\Delta_{\g{m}}$ is the complement of  $\Delta_{\g{k}}$
in  $\Delta$. A  decomposition 
of  $\Delta_{\g{m}}$ in terms of positive and negative root 
subspaces is  compatible 
with the decomposition of $\g{g}=\g{m}\oplus \g{k}$ if  \cite{roemer}
\bqali
\Delta^+_{\g{m}}\cup \Delta_{\g{m}}^-= \Delta_{\g{m}}\, , \ 
\Delta_{\g{m}}^+ 
\cap \Delta_{\g{m}}^- = \emptyset \label{pos1} \ , \\
\alpha , \  \beta \in \Delta_{\g{m}}^+\, , \  \alpha + \beta \in 
\Delta_{\g{m}} 
\ \Rightarrow  \ \  \alpha + \beta \in \Delta_{\g{m}}^+ \label{pos2} \
, \\
\alpha \in \Delta_{\g{m}}^+\, , \  \ \beta \in  \Delta_{\g{k}}\, , \ 
 \alpha + \beta \in
\Delta \ \Rightarrow  \ \alpha + \beta \in \Delta_{\g{m}}^+ 
\label{pos3}  \ . 
\eqali
 The first two conditions above are straightforward to understand,
whereas the last condition implies the $\mathrm{ad}(K)$-invariance of
$\Delta_{\g{m}}^+$.   The choice of a
positive root system in 
$\Delta_{\g{m}}$ is in one-to-one
correspondence with the choice of a Weyl chamber in $\g{h_m}$ \cite{roemer}.

Let  $\{t_m; m=1,2 \ldots ,  \mathrm{dim}(M)\}$ and  
$\{t_a; a=1,2 \ldots , \mathrm{dim}(K)\}$ be a basis 
in $\g{m}$ and $\g{k}$, respectively.
We write 
\bq
g^{-1} \mathrm{d} g = e^m t_m + \omega^a t_a \ ,
\label{leftinv}
\eq
 where  $\{e^m\}$ is the 
frame on $G/K$, $\{\omega^a\}$ is the canonical connection 
on $G/K$ \cite[p. 189]{ nom}, \cite{nom1} and  $g\in G$. 
Then  \cite[p. 193]{nom}
\bqali
&&H^l\!:= \mathrm{D} e^l =  \mathrm{d} e^l + \omega^a \wedge e^m {f_{am}}^l = -
\frac{1}{2!} {f_{mn}}^l e^m \wedge e^n  \label{cartan} \ , \\
&&F^a\! := \mathrm{D} w^a =  \mathrm{d} \omega^a + \frac{1}{2} 
\omega^b \wedge \omega^c 
{f_{bc}}^a = -\frac{1}{2!} {f_{mn}}^a e^m \wedge e^n \label{curva}\ 
\eqali
 are the torsion and the curvature of $G/K$, respectively, where 
$\mathrm{D}$ is the covariant
derivative on
$M$ with respect to the canonical connection $\omega$, 
$[t_m, t_n] = {f_{mn}}^l t_l +
{f_{mn}}^a t_a$ and similarly for the rest of the structure constants. 
Using the canonical normal metric on $\g{m}$, we find that $f_{mnl}$
is totally anti-symmetric and hence
\bq
H= -\frac{1}{3!} f_{lmn}\, e^l \wedge e^m \wedge e^n 
\label{torsion}
\eq
 is a left-invariant three-form\footnote{Our convention
for p-forms is $w_p :=
\frac{1}{p!} w_{[a_1, \ldots , a_p]} e^{a_1} 
\wedge \ldots \wedge e^{a_p}$.} on $G/K$. In addition, 
\bq
\mathrm{d} H = - \mathrm{tr}( F \wedge F)=   -\frac{1}{4}\,
f_{[lm}{}^a f_{no]a} e^l \wedge e^m \wedge e^n \wedge  e^o \ .
\label{dh}
\eq
The homogeneous  tensors on $G/K$ are determined 
by their values on
$T_{eK}(G/K)$ which we have identified with $\g{m}$ 
\cite[p. 193 and 201]{nom}; $e$
is the identity in $G$. In particular, an endomorphism $I$ 
of $\g{m}$ with the properties
\bqali
&&\!\!\!\!\!\!\!\!\!\!\!\!\!\! I^2(X)=-X \label{square}\ ,\\
&&\!\!\!\!\!\!\!\!\!\!\!\!\!\! {[I(X),I(Y)]_{\g{m}}} - 
[X,Y]_{\g{m}} - I([I(X),Y]_{\g{m}}) - 
I([X,I(Y)]_{\g{m}})=0\ , \ \ 
\label{nien}\\
&&\!\!\!\!\!\!\!\!\!\!\!\!\!\! I([Z,X]_{\g{m}})=
[Z,I(X)]_{\g{m}} \label{inva} \ 
\eqali
 induces a homogeneous complex structure on $G/K$, where 
$X,Y \in \g{m}$, $Z \in \g{k}$
and
$[\cdot,\cdot]_{\g{m}}$ is the restriction of the commutator on
$\g{g}$ to $\g{m}$.
We remark that   homogeneous tensors on $G/K$ are  covariantly
constant with respect to the canonical connection. 
These tensors are uniquely determined by their restriction in 
$\g{m}$ and
they correspond to 
$\mathrm{ad}(K)$-invariant tensors on $\g{m}$.

Every hermitian manifold $M$ admits a 
connection of the form (\ref{cona})  with respect to which the 
complex structure $I$ is
covariantly constant \cite{yano, obata}. The torsion of this connection
is 
\bq
H_{mno}= \frac{3}{2} I_m{}^{\tilde{m}} 
I_n{}^{\tilde{n}} I_o{}^{\tilde{o}} 
\partial_{[\tilde{m}}I_{\tilde{n}\tilde{o}]} \ .
\label{hercon}
\eq
 In particular, every hermitian homogeneous space admits
a homogeneous KT structure with respect to the canonical connection.


\section{Homogeneous KT  spaces}

We shall  show that all  homogeneous, 
closed, simply connected
complex  spaces, i.e. C spaces \cite{wang}, admit a KT structure.
Let $K$ be a closed and connected subgroup of a compact 
semi-simple group $G$ whose semi-simple
part coincides with the semi-simple part of the centralizer of
a toral subgroup $H_1$ of $G$.
The Lie algebra, $\g{k}$, of $K$
is reductive, i.e.
$\g{k^c}= \g{u^c} + \g{h^c_0}$, where
$\g{u^c}$ and $\g{h^c_0}$ are  semi-simple and Abelian ideals, respectively.
We  decompose  the Cartan
sub-algebra $\g{h^c}$ of $\g{g^c}$ as
\bq
\g{h^c}=\{\underbrace{h_1,\ldots,h_{a_1}}_{\g{h_m^c}},
\underbrace{\overbrace{h_{a_1+1},\ldots,h_{a_2}}^{\g{h^c_0}},
\overbrace{h_{a_2+1}\ldots,h_{a_3}}^{\g{h_u^c}}}_{\g{h^c_k}}\}\ ,
\label{hdecom}
\eq
 where $\g{h_u^c}$ is the Cartan sub-algebra of
$\g{u^c}$, $1 \leq a_1 \leq a_2 \leq a_3=\mathrm{rank}(\g{g})$ and 
\bq
\g{h^c_1}:=\g{h_m^c} + \g{h^c_0} \  
\label{h1}
\eq
 is the Lie algebra of the toral subgroup $H_1$ of $G$.
It is clear that
\bq
\Delta_{\g{k}}=\{\alpha \in \Delta: \alpha(h)=0 \ ;  \forall h \in  
\g{h^c_1}\} \ . 
\label{dkcom}
\eq
  Relations similar to
(\ref{hdecom}) and (\ref{h1}) hold also for
the associated real Lie algebras and Lie groups. 

An M space is 
a homogeneous space of the form 
$G_s/K$, where $G_s$ is  a compact, simply connected
simple Lie group and the subgroup $K$ is the semi-simple
part of the centraliser of a toral subgroup $T\subset G_s$.  
Since any toral subgroup of $G_s$ is contained in a maximal torus,
M spaces correspond to  decompositions of the form 
(\ref{hdecom}) for which $G=G_s$ is simple and  $\g{h^c_0}=\{0\}$.
The Lie algebra of the torus $T$ is $\g{h_1}$.
All C spaces are fibre decomposition spaces of a product of
M spaces with a torus as fibre.
Thus we have the following classification of  C spaces in terms of
M spaces \cite{wang}:

\begin{theorem}
Let  $G$ be a simply connected, semi-simple, compact 
group with a decomposition as in (\ref{hdecom}).
There is a one-to-one correspondence between C spaces,  $G/K$, and
even-dimensional spaces of the form
\bq
\left(M^1 \times \ldots \times M^r\right)/H_0 \ ,
\label{cdecom}
\eq
  where the $(M^i)$'s, $\{1 \leq i \leq r\}$,   are M spaces
and $H_0$  is a toral subgroup of $G$  with Lie algebra $\g{h_0}$.
\label{theorem0}
\end{theorem}

\no So far we have only
considered simply connected spaces. But a version of Theorem \ref{theorem0}
also holds for spaces with  finite fundamental
group. 

In what follows we shall consider even-dimensional
homogeneous $G/K$ spaces of the form
\bq
\left(M^0 \times M^1 \times \ldots \times M^r\right)/H_0 \ ,
\label{newdecom}
\eq
 where all the $M^i$'s for $1\leq i\leq r$ are M spaces as above,
$M^0$ is a toral group and $H_0$ is a toral subgroup of $G$.
Choosing an invariant metric on $G$, we can arrange for
the decomposition $\g{g}=\g{k}\oplus \g{m}$ to be orthogonal.
Using this, we shall show that all the above spaces are KT manifolds. 

All M spaces can be explicitly constructed from 
the Dynkin diagrams of simple groups.
There are   
four infinite series of Lie algebras,
$A_r, B_r, C_r$ and $D_r$ for 
$\{r= 1,2, \ldots\}$, and  five
exceptional Lie algebras, $E_6,E_7, E_8, F_4$ and $G_2$. 
To identify the sub-algebra  $\g{k^c}$ of  $\g{g^c}$, one takes the
Dynkin diagram of $\g{g^c}$ and colours 
a subset of its vertices. The, possibly disconnected,
sub-diagram consisting of the coloured vertices and the connecting lines
between them is  itself a Dynkin diagram of 
$\g{k^c}$. Then $\g{m^c}$ is defined using (\ref{perp}) and 
(\ref{dkcom}). 
We remark, that the above  embeddings of $\g{k^c}$ in  $\g{g^c}$ 
are regular \cite{dynk}, i.e. 
\bq
\Delta_{\g{k}} \subseteq \Delta \ , \ \ 
\g{h^c_k} \subseteq   \g{h^c} \ .
\eq

To illustrate the construction, we take  $\g{g}\, =E_8$ and
colour its Dynkin diagram as follows:

$$
\epsfsize=0.43\textwidth
\centerline{\epsfbox{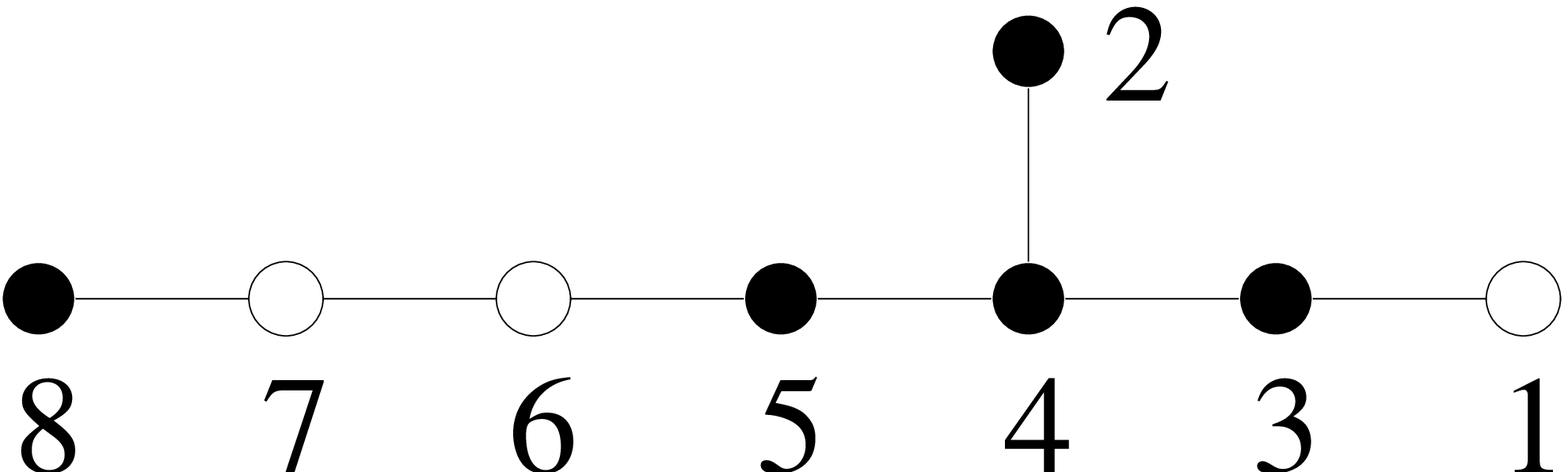}}
$$
\vspace{12mm}
\centerline{Fig. 1: A coloured  Dynkin diagram of $E_8$}

 Then $\g{k^c}$ is $D_4 \oplus A_1$ and  $\g{h_1^c}$ is spanned by
\bqali
&&h_1=2h_{\alpha_1}+h_{\alpha_3}- h_{\alpha_5}-2 h_{\alpha_6} \label{e8a}\ , \\
&&h_2=-2h_{\alpha_1}+ h_{\alpha_2}+ 2 h_{\alpha_4}+3  h_{\alpha_5}+4
h_{\alpha_6} \ , \\ 
&&h_3=2 h_{\alpha_7}+ h_{\alpha_8} \ , 
\label{e8b}
\eqali
 where the ordering of the simple roots is shown in Fig. 1. It is easy
to check that the semi-simple part of the centralizer of the 
generators $h_1$, $h_2$ and $h_3$ in  $E_8$ is indeed the sub-algebra 
$D_4 \oplus A_1$. The M space associated to 
this decomposition is $E_8/(SO(8) \times SU(2))$,
which is 217-dimensional. This is not a complex space,
but 
\bq
\frac{E_8\times U(1)^b}{SO(8) \times SU(2)\times U(1)^a} 
\label{exam}
\eq
  are complex  spaces for $a=\{0,1,2,3\}$
and $b=\{1,0,1,0\}$.  For the homogeneous spaces 
(\ref{exam}), the Lie algebra of $U(1)^a$ is spanned by a linear
combination of the generators (\ref{e8a}) - (\ref{e8b}) and the
generator  of $U(1)^b$.

To show that all  
homogeneous, complex compact spaces admit  KT structures, we introduce
an endomorphism $I$ on $\g{m^c}$. This can be 
done in two steps. First we define the action of $I$ on
the step operators  $\{e_{\alpha}\}$ in $\g{m^c}$  as
\bq
I(e_{\alpha})=i\epsilon_{\alpha}e_{\alpha} \ , 
\label{comstep}
\eq
  where $\epsilon_{\alpha}$ are real constants.
The requirement that  $I^2=-1$ (\ref{square}) and  
that  $I$  in $\g{m^c}$ is induced by an endomorphism in
 $\g{m}$ constrains the constants  $\epsilon_{\alpha}$ to be
\bq
\epsilon_{ \alpha} \in \{1,-1\} \,  , \ \ 
\epsilon_{+ \alpha} = - \epsilon_{- \alpha} \label{what} \ . 
\eq
 The invariance of $I$ 
implies that
\bq
\alpha \in \Delta_{\g{m}} \, , \ \beta \in \Delta_{\g{k}} \, , \ \alpha +
\beta \in \Delta_{\g{m}} \ \Rightarrow
\epsilon_{\alpha}=\epsilon_{\alpha+\beta} \ .
\label{what1}
\eq
 Finally, the  integrability
condition  (\ref{nien})  for $I$ leads to
\bq
\alpha, \beta \in \Delta_{\g{m}}\, , \ \alpha+\beta \in 
\Delta_{\g{m}} \, , \ 
\epsilon_{\alpha}=\epsilon_{\beta} \ \Rightarrow  
\epsilon_{\alpha}=\epsilon_{\beta}=\epsilon_{\alpha+\beta} \ . 
\label{what2}
\eq
 Comparing the relations (\ref{what}) - (\ref{what2}) with 
the relations (\ref{pos1}) - (\ref{pos3}), we note that they have the
same structure. Thus
there is a one-to-one correspondence between the set of Weyl
chambers of $\g{h_m^c}$ and the set of constants $\{\epsilon_{\alpha}\}$,
which define $I$ on  the set of step operators 
$\{e_{\alpha}\}$ in $\g{m^c}$. This
correspondence is obvious if we  assign to each $\epsilon_{\alpha}$ a
value $\pm 1$, depending on whether the root $\alpha$ is
positive or negative with respect to a chosen Weyl chamber.
We note that the choice of a Weyl chamber for the
definition of the positive roots 
is independent of the choice of a Weyl chamber for the
definition of the endomorphism (\ref{comstep}).  

It remains to define the endomorphism $I$ in the  
Cartan sub-algebra $\g{h_m^c}$.
The dimension of $\g{h_m^c}$ is even by assumption, i.e. 
$\mathrm{dim}(\g{h^c_m})= 2\, \ell$. We begin by taking a
basis in $\g{h_m^c}$  and mapping  it to the $2\,\ell$
generators of $\bR^{2\ell}$, say $\{h_i, h_{\ell+i}:1 \leq i\leq \ell \}$.
There is a $4\,\ell^2$ parameter freedom in doing this. 
Then we define  $I$ on  $\g{h_m^c}$ as
\bq
I(h_i)= h_{\ell+i} \, , \ \ I(h_{\ell+i})= - h_i \ .
\label{comstr}
\eq
 One can verify that the endomorphisms (\ref{comstep}) and
(\ref{comstr}) induce homogeneous complex structures on $\g{m}$.

The conditions for the 
hermiticity of the canonical normal  metric  with respect to
the complex structures (\ref{comstep}) and (\ref{comstr})
are
\bqali
&&B(E^+_{\alpha}, E^+_{\alpha}) = B(E^-_{\alpha}, E^-_{\alpha}) \ ,
\label{herm1}\\ 
&& B(H_i,H_j)=  B(H_{\ell+i},H_{\ell +j}) \label{herm2} \ , \\
&&B(H_i,H_{\ell +j})= - B(H_j ,H_{\ell +i}) \label{herm3} \ . 
\eqali
 Comparing  (\ref{herm1}) with (\ref{killcom1}) we see that
the first  condition is automatically satisfied, whereas the conditions 
(\ref{herm2})  and (\ref{herm3}) place constraints on the
metric restricted to the Cartan sub-algebra. There are
at most $\ell(\ell +1)$ constraints, which  can be 
satisfied by tuning the same number  of
parameters in the de\-fi\-ni\-tion of the complex
structure (\ref{comstr}). This leads to a family of complex structures  
which are compatible with the 
canonical normal metric and have  at least $\ell(3\ell-1)$
parameters:

\begin{theorem}
All  homogeneous spaces of the form 
$(M^0 \times M^1 \times \ldots \times M^r)/H_0$ (\ref{newdecom}) 
admit infinitely many  KT structures.
\label{theoremc}
\end{theorem}
We remark that in particular all compact, even-dimensional  
group spaces with a reductive Lie algebra are KT spaces.

In the construction of KT spaces employing Dynkin diagrams the question arises 
whether differently coloured Dynkin diagrams of
a simple algebra $\g{g}$ lead to equivalent M spaces.
It turns out that 
Dynkin diagrams which are related by (i) outer 
automorphisms and (ii)
special Weyl transformations of $G$,
which preserve the notation of
positivity in $\g{m}$, should be considered equivalent.
This result is similar to that found in \cite{roemer} for K spaces. 
A list of M spaces, which takes the
equivalence of differently coloured Dynkin diagrams into account,
can be found in \cite[Tables 3 and 4]{roemer}\footnote{The original table of
M spaces by Wang \cite{wang} is not complete.}.  They are really 
tables  of simple homogeneous K spaces. But since these
homogeneous K spaces
and the   homogeneous KT spaces have the
same complex structures on the space of step operators, they
are also lists of M spaces with a KT structure.
Note that these  K and KT spaces are equipped with different metrics.
In the case of KT spaces the
normal canonical metric  $B$ is used, whereas in the 
case of K spaces the metric $\tilde{B}$ is constructed from the
symplectic structure
on the co-adjoint orbit, the Kirillov-Kostant-Souriau form, 
and the complex structure.
Only in the case of KT spaces with  $H \subset K$, where $H$ is a
maximal torus of $G$, it is possible to define
non-compact duals for  compact KT spaces in the same way 
as for compact 
K spaces \cite{roemer}. It is interesting to
note that a homogeneous  space $M$, which is K\"{a}hler
with respect to the metric
$\tilde{B}$, also admits a   KT structure  with respect to the
metric $B$ and with $B=\tilde{B}$ if and only if  $M$ is symmetric.


\section{ Homogeneous HKT spaces}

Our main task in this section is to find the subclass of 
the homogeneous KT spaces  which admit HKT structures.  
In order to do this we shall use a decomposition of the group spaces
$G\times^{{\!\!\!\!\!}^{m}} U(1)$, $m \in \bN$, where $G$ is
a semi-simple Lie group with Lie algebra $\g{g}$. Let us set
$\g{g}^{\g{c}}_1=\g{g^c}$ and $\Delta_1=\Delta$. We  
choose a  highest root  $\psi_1$  
in  $\Delta_1$ and define the
three-dimensional complex sub-algebra isomorphic to $\g{sl(2)^c}$ as
$\g{d}^{\g{c}}_1= \mathrm{span}\{ e_{\psi_1},  e_{-\psi_1}, h_{\psi_1}\}$.
Next, we find the
centralizer $\g{b}^{\g{c}}_1$ of $\g{d}^{\g{c}}_1$ in
$\g{g}^{\g{c}}_1$ and define the
compliment $\g{f}^{\g{c}}_1$ of
$\g{b}^{\g{c}}_1 \oplus \g{d}^{\g{c}}_1$ in $\g{g}^{\g{c}}_1$. 
Then the  first  level of the 
decomposition of  $\g{g}^{\g{c}}$ is 
\bq
\g{g}^{\g{c}}_1 =\g{b}^{\g{c}}_1 \oplus 
\g{d}^{\g{c}}_1 \oplus \g{f}^{\g{c}}_1 \ .
\label{deone}
\eq 
  For the second level of the decomposition of $\g{g^c}$, we set  
$\g{g}^{\g{c}}_2\! :=\g{b}^{\g{c}}_1$ and decompose $\g{g}^{\g{c}}_2$ 
using the same  procedure as for $\g{g}^{\g{c}}_1$
in the first level, and so on. This process of 
decomposing $\g{g^c_k}$'s
is continued until a $\g{b}\! :=\g{b}_n$ is found which is
Abelian. This indicates that the decomposition of $\g{g^c}$  
is completed. In the same way it is possible to  decompose 
the associated compact, real Lie algebra $\g{g}$.
In this case,
the three-dimensional sub-algebras $\g{d}_k$ 
are isomorphic to $\g{su(2)}$. Thus a reductive Lie algebra $\g{g}$
can be decomposed as \cite{joy}
\bq
\g{g}= \g{b}   \, 
\oplus^{{\!\!\!\!\!}^{n}}_{{\!\!\!\!\!\!\!\!}_{_{k=1}}} \! \g{d}_k  
 \, \oplus^{{\!\!\!\!\!}^{n}}_{{\!\!\!\!\!\!\!\!}_{_{k=1}}} \! \g{f}_k \ , 
\label{decom}
\eq
 where $\g{b}$ is Abelian, the  $\g{d}_k$'s are highest root  subspaces 
isomorphic to $\g{su(2)}$ and 
the $\g{f}_k$'s are (possibly empty) subspaces of $\g{g}$. We remark
that  the Cartan sub-algebra of $\g{g}$ is contained in 
$\g{b} \oplus_{{\!\!\!\!\!\!\!\!}_{_{k=1}}}^{{\!\!\!\!\!}^{n}}  \g{d}_k$.

To define a hyper-complex structure on $G \times^{{\!\!\!\!\!}^{m}} U(1)$, we
take an integer $m$ such that the dimension of 
$\g{b}\,  \oplus^{{\!\!\!\!\!}^{m}}\, \g{u(1)}$
is equal to the maximal level $n$  in the decomposition of $\g{g}$.
Then we choose a basis $\{ U_k;k=1,2, \ldots ,n\}$ in 
$\g{b}  \oplus^{{\!\!\!\!\!}^{m}}
\g{u(1)}$ and define an isomorphism $\phi_k: \g{su(2)} \mapsto  \g{d}_k$.
The endomorphisms $I_r$ on $\g{b}   \, 
\oplus^{{\!\!\!\!\!}^{n}}_{{\!\!\!\!\!\!\!\!}_{_{k=1}}} \! (\g{d}_k  
\oplus \g{u(1)}_k)$ are 
\bqali
&&I_r(U_k)=\phi_k(Y_r) \ , 
\label{comstr1}\\
&&I_r(\phi_k(Y_s))= - \delta_{rs} U_k + \epsilon_{rs}{}^t \phi_k(Y_t) 
\label{comstr2} \ , 
\eqali
 where $\{Y_1,Y_2, Y_3\}$ is a  basis in $\g{su(2)}$, and  
on $\g{f}_k$ are 
\bq
I_r(E^{\pm}_{\beta})= [E^{\pm} _{\beta}, \phi_k(Y_r)] \ ,
\label{comstr3}
\eq
 where the $E^{\pm}_{\beta}$'s are  step operators in 
$\g{f}_k$. These endomorphisms preserve the levels of
the decomposition $\g{g}$ and depend on the choice of the basis
in $\g{b}  \oplus^{{\!\!\!\!\!}^{m}}
\g{u(1)}$, i.e. there are $n^2$ inequivalent choices of $I_r$ 
in $\g{g}\oplus^{{\!\!\!\!\!}^{m}}
\g{u(1)}$.  We remark that  different
choices of the isomorphisms $\phi_k$ are related to different choices
of the constants $\epsilon_\alpha$ in the definition of 
the complex structure
(\ref{what}).  

To investigate the  endomorphisms $I_r$ in more detail, we
take without loss of generality
$(\phi_k(Y_1),\phi_k(Y_2),\phi_k(Y_3)) = (H_{\psi_k}, E_{\psi_k}^+, E_{\psi_k}^-)$.
On each subspace
$\g{d}_k \oplus \g{u(1)}_k$,  $1 \leq
k \leq n$,  the endomorphism $I_1$ is defined as 
\bq
I_1(E^{\pm}_{\psi_k})= \pm E^{\mp}_{\psi_k}\, , \ 
I_1(U_k)= H_{\psi_k} \, ,    \  I_1(H_{\psi_k})= - U_k \  
\eq
 and on  $\g{f}_k$  as
\bq
I_1(E^{\pm}_{\beta})=\mp  2 \frac{\psi_k \cdot \beta}{\psi_k\cdot \psi_k}
E^{\mp}_{\beta} \ . 
\eq
 We remark that if we set $h_{k}:= i H_{\psi_k}$, 
$h_{\ell+k}:= i U_k$ and $n=\ell$, then 
$I_1$ is the same as  $I$  given in (\ref{comstep}) and
(\ref{comstr}). However the other two endomorphisms, $I_2$ and $I_3$, are
not of the same form  because they 
interchange step operators with
Cartan sub-algebra generators. On each subspace
$\g{d}_k \oplus \g{u(1)}_k$, $1 \leq
k \leq n$,  the endomorphism $I_2$ is defined as 
\bq
I_2(E^-_{\psi_k})= H_{\psi_k} \, ,  \ I_2(H_{\psi_k})= -E^-_{\psi_k} \, ,  \
I_2(U_k)=  E^+_{\psi_k}\, ,  \ I_2(E^+_{\psi_k})= -U_k 
\label{i2a}\  
\eq
 and on $\g{f}_k$  as
\bq
I_2(E^{\pm}_{\beta})=  N_{\psi_k,-\beta} E^{\mp}_{\psi_k-\beta} \, , \
\ 
I_2(E^{\pm}_{\psi_k-\beta})= - N_{\psi_k,-\beta}E^{\mp}_{\beta} 
\label{i2d} \ .
\eq
 Note that 
if   $E_{\beta}^{\pm}$ is in  $\g{f}_k$, then
$E_{\psi_k-\beta}^{\pm}$ is also in $\g{f}_k$. We
can assume without loss of generality 
that $ N_{\psi_k,-\beta}$ is positive, since if it is
negative we can exchange  $\beta$ with $\psi_k - \beta$.
The endomorphism $I_3$ can be expressed in a similar way.

The endomorphisms $I_r$ equip $\g{g}\oplus^{{\!\!\!\!\!}^{m}}
\g{u(1)}$ with a hyper-complex structure provided
that 
\bq
(N_{\psi_k, - \beta})^2 =1 \, , \ \ \ 2 \, 
\frac{\psi_k \cdot \beta}{\psi_k \cdot \psi_k}=+1 
\label{cond}
\eq
 for $1 \leq k \leq n$, roots $\beta$ with
$E_{\beta}^{\pm} \in f_k$ and  highest
roots $\psi_k$ in $\Delta_k$. 
The conditions (\ref{cond}) are  satisfied in the basis
(\ref{newcom1}) - (\ref{newcom3}). The
first follows from  the definition of $\psi_k$ as a highest root
in $\Delta_k$ 
and  relation (\ref{stru}), the second is derived using
the Bianchi identity  
$[[e_{\psi_k},e_{-\psi_k}], e_{-\beta}] +{\rm cyclic}=0$.
The vanishing of the  Nijenhuis tensors can be verified by 
direct computation, as in the last section, or by using an 
argument by  Samelson
\cite{sam} as generalized in \cite{joy}. The hyper-complex structures are 
homogeneous by construction. 

The  hyper-complex structures on group spaces 
$G\times^{{\!\!\!\!\!}^{\tilde{m}}} U(1)$  induce hyper-complex structures 
on homogeneous 
spaces $(G/K) \times^{{\!\!\!\!\!}^{m}} U(1)$ using the fact
that the hyper-complex structures on 
$G\times^{{\!\!\!\!\!}^{\tilde{m}}} U(1)$ 
are constructed for every level of the decomposition (\ref{deone})
separately.  Thus  we decompose $\g{g}$ as before but instead of
continuing  until we find a $\g{b}_n$ which 
is Abelian, we stop 
at some level $l$, $1 \leq l\leq n$. 
Let us denote the semi-simple part of  $\g{b}_l$ as
$\g{\tilde{b}}_l$. The Lie algebra 
$\g{k}$ comprises $\g{\tilde{b}}_l$
and possibly some of the Abelian ideal of $\g{b}_l$, i.e. 
$\g{\tilde{b}}_l \subseteq \g{k} \subseteq \g{b}_l$.
The integer $m$ is  fixed by the condition that the dimension of 
$(\g{b}_l/ \g{k})  \oplus^{{\!\!\!\!\!}^{m}} \g{u(1)}$ is  equal 
to the level $l$ of the decomposition.
This proves  \cite{joy}:

\begin{theorem}
Let $G$ be  any closed, semi-simple, simply connected, compact
Lie group,  $G_l$ a
subgroup of  $G$ as defined above 
for  some integer  $l$,  $1 \leq l \leq n$,  
$\tilde{B}_l$ the semi-simple part of  $B_l$ 
and $K$ a
subgroup of $G$ with  $\tilde{B}_l \subseteq
K \subseteq B_l$, 
then there exists an integer $m$ with
$0 \leq  m \leq l$, such that $(G / K)  
\times^{{\!\!\!\!\!}^{m}} U(1)$ admits infinitely many  hyper-complex
structures.
\label{theorem2}
\end{theorem}

Note that all group manifolds which admit hyper-complex structures 
\cite{spin}, \cite{joy} are included in Theorem 
\ref{theorem2} for $K =\tilde{B}_l = \{ 1\}$.
Note also that every Abelian group of dimension $4\,d$ is naturally
hyper-complex. To see this one maps the $4\,d$ $\g{u(1)}$-generators into 
$\bH^d$, where $\bH$ are the quaternions. There is a freedom in
doing this of $4\,d$ parameters. Then the action of the
hyper-complex structure on $\bH^d$  is generated by the natural 
action of the quaternions. Thus if $(G/K)\times^{{\!\!\!\!\!}^{m}}
U(1)$ is hyper-complex, then $(G/K)\ \times^{{\!\!\!\!\!\!\!\!\!}^{m+4d}}
U(1)$ is hyper-complex as well. 

A more general class of HKT spaces can
be constructed which are of the form
$(G\ \, \times^{{\!\!\!\!\!\!\!\!\!}^{m+4d}}U(1))/K$. It is understood
that taking the coset with $K$  mixes the $U(1)$ generators with the
generators of $H$ that lie in $B_l$. It follows that 
the resulting HKT space may not be of the form of Theorem \ref{theorem2}.

To show that these hyper-complex homogeneous spaces  admit  HKT
structures, we consider  the
conditions under which $B$  is tri-hermitian. It turns out, that 
the conditions for  $B$ to be tri-hermitian on the subspace
$\oplus^{{\!\!\!\!}^{l}}_{{\!\!\!\!\!\!\!\!}_{_{k=1}}} \! 
\g{f}_k$ are the same as those of (\ref{cond}).
The conditions for $B$ to be tri-hermitian on
the subspace $\oplus^{{\!\!\!\!}^{l}}_{{\!\!\!\!\!\!\!\!}_{_{k=1}}} \! 
(\g{d}_k\oplus \g{u}_k)$ are
\bq
B(E_{\psi_j}^+,E_{\psi_k}^+)=B(E_{\psi_j}^-,E_{\psi_k}^-)= 
B(H_{\psi_j},H_{\psi_k}) = B(U_{j },U_{k}) \label{hermi1}\ 
\eq
  for $1 \leq j, k \leq l$; all other components of $B$ must vanish.
Using (\ref{killcom1}) we find
\bq
B(E_{\psi_j}^+,E_{\psi_k}^+)=B(E_{\psi_j}^-,E_{\psi_k}^-)=
\frac{4}{\psi_j \cdot \psi_j}\, \delta_{j,k}  \label{hermi2}\ . 
\eq
 In the next section it will become apparent, that the generators
$H_{\psi_j}$ are mutually orthogonal by construction.  The
generators $E^{\pm}_{\psi_j}$ and $H_{\psi_j}$ are of equal  length for each  
$j=1,2,\ldots ,l$ because they span  a standard
basis of $\g{su(2)}$, for which this is true. Thus it is left to
determine the Cartan sub-algebra generators $U_j$, such that 
\bq
B(U_j,U_k)= \frac{4}{\psi_j \cdot \psi_j}\, \delta_{j,k}  
\label{hermi3}\ . 
\eq
 This imposes at most $\frac{l(l+1)}{2}$ constraints 
on the complex structures,
which leave at least $\frac{l(l-1)}{2}$  inequivalent 
HKT structures, constructed from the $l^2$ hyper-complex structures:

\begin{theorem}

The homogeneous manifolds $(G/K) \times^{{\!\!\!\!\!}^{m}} U(1)$ 
of  Theorem \ref{theorem2} with more than one level of decomposition
(\ref{decom}) admit infinitely many  
inequivalent HKT  structures, and finitely many otherwise.
\label{theorem3}
\end{theorem}


\section{Homogeneous  HKT  spaces and ex\-ten\-ded Dyn\-kin diagrams}

The decomposition (\ref{decom}) can be most easily described using
extended Dynkin diagrams. This is so  because the addition
of the highest root to the extended  Dynkin diagram introduces 
a natural colouring in the standard Dynkin diagram, which 
is associated with
homogeneous HKT spaces. The use of Dynkin diagrams also
enables  us to compile  lists of  homogeneous HKT spaces.

The extended Dynkin diagram of  a simple Lie algebra $\g{g}$ is
found by adding one  vertex  to 
the Dynkin diagram of $\g{g}$ which represents the highest root. 
A list of the extended Dynkin diagrams of  
simple Lie algebras is given in
Table 1 (see e.g. \cite{fuchs}). The simple roots 
$\{\alpha_i\}$ are marked by  coloured and
uncoloured  vertices  whereas the highest
root $\psi$ is marked by a crossed vertex. In Table 1, we  also 
include the  symmetry groups $\bar{\Gamma}$
of the uncoloured  extended Dynkin  diagrams.
Coloured Dynkin diagrams which are related by outer automorphisms
of the uncoloured  extended Dynkin diagram lead to equivalent HKT structures. 

$$
\epsfsize=0.99\textwidth
\centerline{\epsfbox{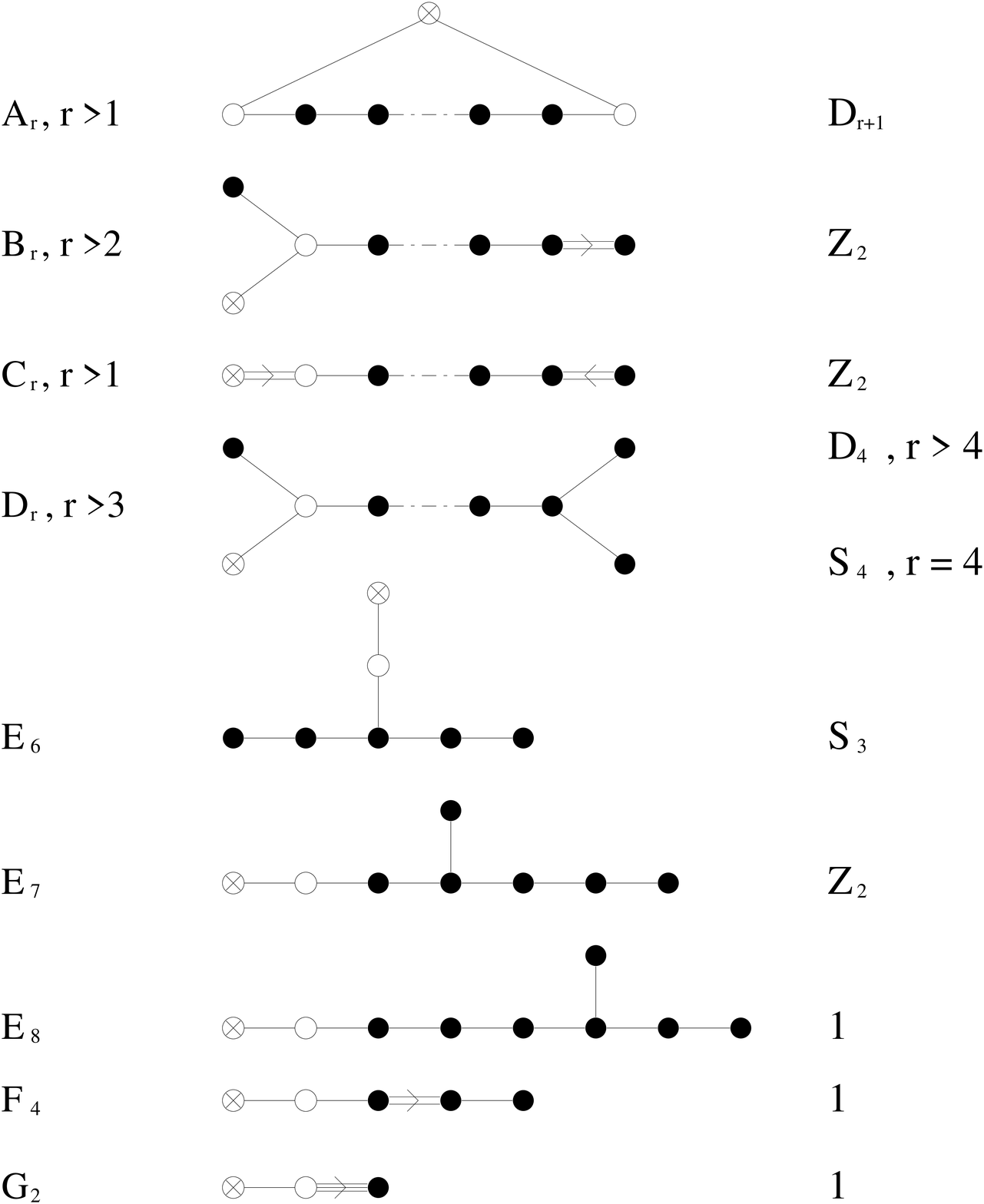}}
\label{dynkin}
$$
\vspace{3mm}
\begin{center}
Table 1: The decomposition (\ref{deone}) of the extended Dynkin diagrams
of simple Lie groups and their symmetry group $\bar\Gamma$.
\end{center}

The extended Dynkin diagram of
$A_r$ has the property that the highest root is connected to two simple roots
and its decomposition is different from that
of the other simple Lie algebras. So it will be 
treated separately. The decomposition
of the remaining simple Lie algebras 
involves the following steps:

{$\bullet$} In the first level we  set $\g{g}_1\! :=\g{g}$
and colour  
the extended  Dynkin diagram as in Table 1.
The highest root subspace 
$\g{d}_1$ is isomorphic to $\g{su(2)}$ spanned by 
$\{ E^{\pm}_{\psi_1},  H_{\psi_1}\}$. 

{$\bullet$} Using (\ref{newcom1}) - (\ref{newcom3}), we find that  
the centralizer, $\g{b}_1$, of $\g{d}_1$ in $\g{g}_1$ is
isomorphic to the sub-algebra associated with the coloured 
sub-diagram of the extended Dynkin diagram.

{$\bullet$} The
subspace $\g{f}_1$ consists of all step operators
$E^{\pm}_{\alpha}$ 
of roots $\alpha$, other than $\psi_1$,  
which have a non-zero expansion
coefficient  associated with the simple root marked with an  uncoloured vertex
in the diagram. 

{$\bullet$} Next we set  $\g{g}_2\! :=\g{b}_1$ and 
distinguish two cases, depending on whether 
$\g{g}_2$ is simple or not. In the former case, we use
the extended Dynkin diagram
of $\g{g}_2$ from Table 1 and repeat the decomposition as  
for  $\g{g}_1$ above. If
$\g{g}_2$  is not simple,  it is always of the form $A_1\oplus
\g{\tilde{g}}_2$, where
$\g{\tilde{g}}_2$ is simple. We can proceed  by taking  the 
highest root  $\psi_2$ of
$\g{g}_2$  either in $A_1$ or  in  $\g{\tilde{g}}_2$. In the former  
case   we find
$\g{d}_2=A_1$, 
$\g{f}_2=\emptyset$ and  $\g{b}_2=\g{\tilde{g}}_2$.
In the latter case, we use the   extended Dynkin diagram of 
$\g{\tilde{g}}_2$ from Table
1 and repeat the decomposition as for  $\g{g}_1$ above. 

This process is continued until 
either a $\g{b}_{n}$ is found which is Abelian or a $\g{b}_k$  
is found which is equal to
$A_r$ for 
$r>1$. So it remains to describe the decomposition of the Lie 
algebra $A_r$ for  $r>1$
using Dynkin diagrams.  For this we use the following steps:

{$\bullet$} In the first level we set  
$\g{g}_1:=\g{g}$ and identify the highest root subspace $\g{d}_1$ 
as above. Then
we colour  the
extended Dynkin diagram of  $A_r$ as in Table 1. 

{$\bullet$} The  centralizer, $\g{b}_1$, of $\g{d}_1$ in $\g{g}_1$
is equal to  $ A_{r-2}  +
\g{u(1)}$, where the $\g{u(1)}$-generator 
commutes with the highest root step operators
$E^{\pm}_{\psi_1}$ and satisfies a certain orthogonality
condition which we shall explain later. The coloured part of the  Dynkin
diagram  is that of a $A_{r-2}$ sub-algebra.
 
{$\bullet$} The subspace $\g{f}_1$ consists of all 
step operators $E^{\pm}_{\alpha}$ for roots $\alpha$,   
other than $\psi_1$, 
which have  non-zero expansion
coefficients  associated with at least one of the two simple roots 
marked with an  uncoloured
vertex in the  diagram. 

{$\bullet$} In the second level we set $\g{g}_2=A_{r-2}  + \g{u(1)}$ 
and repeat the
decomposition as above. 

We note that $\g{b}_2$ is
equal to  $ A_{r-4}  + \g{u(1)} + \g{u(1)}$, where the two 
$\g{u(1)}$-generators are linear independent and commute with 
$E^{\pm}_{\psi_1}$ and $E^{\pm}_{\psi_2}$. 
This process is continued until  a 
$\g{b}:=\g{b}_n$ is found which is Abelian. 
The $n= [r/2]$ 
$\g{u(1)}$-generators  in $\g{b}$ commute with all
highest root subspace $\g{d}_k$ for $1 \leq k \leq n$.

Let us consider in detail how to choose the $\g{u(1)}$-generators in
$\g{b}$ arising from the above  decomposition of $A_r$ under the
additional assumption of
the tri-hermiticity of the metric $B$. As we have seen these are the 
additional
conditions necessary to find HKT structures.
The set of roots of $A_r$ can be taken as $\Delta=
\{\epsilon_p - \epsilon_q; 1 \leq p \neq q \leq r+1\}$, the set of
positive roots as  $\Delta^+=
\{\epsilon_p - \epsilon_q; 1 \leq p < q \leq r+1\}$ and the set of
simple roots as
$\Delta^s = \{ \alpha_i=\epsilon_i - \epsilon_{i+1}; 1 \leq i \leq
r\}$, where $\{\epsilon_p; p=1, \dots, r+1\}$
is an orthonormal basis in $\bR^{r+1}$ (for details see \cite[p. 860]{corn}). 
The metric $B$ is
\bq
B(E_{\alpha}^+,E_{\alpha}^+)= B(E_{\alpha}^-,E_{\alpha}^-)=
B(H_{\alpha},H_{\alpha})= \frac{4}{ \alpha \cdot \alpha}= 4(r+1) \ 
\label{killinga}
\eq
 for all $\alpha \in \Delta^+$ and 
with all other components of $B$ vanishing, except 
$B(H_{\alpha_i},H_{\alpha_{i+1}})= -2(r+1)$. For  $r=2n$ the
number of levels in the decomposition (\ref{decom}) is $n$, and
the $n$ highest roots $\psi_k$ are $\epsilon_k - \epsilon_{2n-k}$,
$k=1,2,\ldots,n$. The $n$ $\g{u(1)}$-generators $U_k$ are  constrained
by

\begin{description} 
\item[(i)] $\{ U_k, H_{\psi_k};k=1,2,\ldots,n\}$  span $\g{h}$, 
\item[(ii)] the $U_k$'s commute with  all step operators
associated with highest roots, i.e.  
\bq
[ U_k, E^{\pm}_{\psi_l}]=0
\label{ukcon}
\eq
  for all $k,l=1,2,\ldots,n$ and
\item[(iii)] $B$ is tri-hermitian on the Cartan sub-algebra with respect
to the complex structures of the previous section, i.e. 
\bq
B(U_k,U_l) = 4(2n+1)\delta_{k,l} \,  ,  \ \ 
B(U_k,  H_{\psi_l}) = 0 \ 
\label{condh}
\eq
 for all $k, l= 1,2, \ldots ,n$. We remark that the last condition implies that
the $\g{u(1)}$ generators are orthogonal to $\g{d}_l$. 
\end{description} 

To find a solution to all these conditions we expand 
\bq
U_k= (c_k)^k H_{\alpha_k} +  (c_k)^{k+1} H_{\alpha_k} + \cdots 
+ (c_k)^{2n-k} H_{\alpha_{2n-k}} 
\label{suexp}
\eq
 in terms of Cartan
sub-algebra generators of simple roots which lie in the subspaces $\g{b}_k$, where the 
coefficients are real numbers. Condition (\ref{ukcon}) implies 
\bq
(c_k)^k = -(c_k)^{2n-k} , (c_k)^{k+1} = -(c_k)^{2n-k-1}, \ldots ,
(c_k)^{n}+ = -(c_k)^{n+1} \ ,
\label{restc1}
\eq
 whereas condition (\ref{condh}) determines the  $(c_j)^k$ to be
\bq
(c_k)^l = [2(n-l)+1](c_k)^{n} \, ,  \ \ 
(c_k)^{n} = 
\frac{1}{\sqrt{4(n-k+1)^2-1}} \ 
\label{restc2}
\eq
 for $1 \leq k \leq n$ and  $k \leq l \leq n-1$.
In the next section we shall argue,
that  the  general solution of the $U_k$'s is generated from the
special solution (\ref{restc2}) applying an orthogonal transformation. 
 
A similar analysis is possible for the case of $A_r$
for $r=2n-1$. The  Cartan sub-algebra  splits in 
$n$ highest root generators and $n-1$ generators  $\{U_{\tilde{k}};
\tilde{k}=1,2,\ldots , n-1\}$. The  expansion coefficients of the
$U_{\tilde{k}}$'s are
\bqali
&&\hspace{-15mm} 
(c_{\tilde{k}})^{\tilde{k}} = -(c_{\tilde{k}})^{2n-{\tilde{k}}} , \
(c_{\tilde{k}})^{\tilde{k}+1} = -(c_{\tilde{k}})^{2n-\tilde{k}-1},
\ldots , \
(c_{\tilde{k}})^{n-1}+ = -(c_{\tilde{k}})^{n+1} \ , \label{restc3a}\\  
&&\hspace{-15mm} (c_{\tilde{k}})^{n-1}=0 \ ,  \\ 
&&\hspace{-15mm}(c_{\tilde{k}})^l = [n-l](c_{\tilde{k}})^{n-1} \, , \  \ 
(c_{\tilde{k}})^{n-1} =  \frac{1}{\sqrt{(n-\tilde{k})(n-\tilde{k}+1)}} \ 
\label{restc3b}
\eqali
 for $1 \leq \tilde{k} \leq n-1$ and $\tilde{k} \leq l \leq n-2$. 

As an example, we take $A_r=A_4$.  Implementing 
the above decomposition, we find 
\bqali
&& H_{\psi_1}=H_{\alpha_1}+H_{\alpha_2}+H_{\alpha_3}+H_{\alpha_4}\ , 
\label{su5a}\\
&& H_{\psi_2}=H_{\alpha_2}+H_{\alpha_3} \ , \\
&& U_1=\frac{1}{\sqrt{15}}\, (3H_{\alpha_1}+
H_{\alpha_2}-H_{\alpha_3}-3H_{\alpha_4})\ , \\
&& U_2=\frac{1}{\sqrt{3}}\, (H_{\alpha_2}-H_{\alpha_3}) \ .
\label{su5b}
\eqali

We note, that all  $\g{u(1)}$-generators in
$\g{b}$ arise from the decomposition of a sub-algebra 
$\g{b}_k = A_r$ for some $r > 1$. Thus there are 
two types of reductive group spaces admitting an HKT structure,
those that
have a non-trivial Abelian part $\g{b}$  and
those that have a trivial one. The former  are associated 
with the algebras $A_r, D_{2r
+1}$ and $E_6$ and are of the form 
$G \times^{{\!\!\!\!\!}^{m}} U(1)$ for $0 \leq m 
< \mathrm{rank}(\g{g})$. The later are
associated with  the 
algebras $B_r, C_r,
D_{2r}, E_7, E_8, F_4$ and $G_2$ and are of the form
$G \times^{{\!\!\!\!\!}^{m}} U(1)$ for  $m=\mathrm{rank}(\g{g})$. 

In the previous section to prove the tri-hermiticity of $B$
 we have used the fact that the Cartan 
sub-algebra generators of the highest root spaces
$\g{d}_k$ are mutually orthogonal. 
This is easily verified using Dynkin diagrams  because
the associated crossed  vertices of the highest roots of the different levels 
are unconnected in the extended  Dynkin diagrams. 

Homogeneous HKT spaces associated with reductive Lie
algebras $\g{g}$, as homogeneous KT spaces, are of the
form
\bq
M= \left(M^0 \times M^1 \times \ldots \times M^r\right)/T \ , 
\label{haha}
\eq
 where $M^0$ and $T$ are  toral groups and $M^i$ are homogeneous HKT spaces
associated with simple groups. Let  $\g{g}= \g{b}^0 \oplus  
\g{g}^1\oplus\cdots 
\oplus\g{g}^r$ be a reductive Lie algebra, where 
$\g{b}^0$ is Abelian and  $\g{g}^i$,  $1 \leq i
\leq r$, are simple ideals. The homogeneous space $M$  can be 
constructed in the following steps:  

{$\bullet$} We first decompose 
the simple ideals  $\g{g}^i$ separately and determine the
HKT spaces $M^i$, as in the beginning of the section. 

{$\bullet$} We set  $\g{b}=  \g{b^0}
\oplus^{{\!\!\!\!\!}^{r}}_{{\!\!\!\!\!\!\!\!}_{_{i=1}}} \! \g{b}^i$, 
where $\g{b}^i$ 
is the Abelian part of the decomposition (\ref{decom}) 
of the simple ideal $\g{g}^i$. Then we
divide out some part of $\g{b}$ say $\g{t} \subset \g{b}$. In particular 
$\g{t}$ can be a linear
combination of $\g{u(1)}$-generators of $\g{b}^0$ and 
$\g{b}^i$'s. 

{$\bullet$} Since for the construction of a HKT structure every highest
root subspace is paired with a
$\g{u(1)}$ generator, we add $c$ $\g{u(1)}$ generators such that
$b+c-a$ is zero or positive and
divisible by four, where $a=\mathrm{dim}(\g{t})$ 
and $b=\mathrm{dim}(\g{b}^0)$. Then 
$M^0$ in (\ref{haha}) is spanned by $b+c$ $\g{u(1)}$-generators.  Our
results are summarized as:

\hspace{-1.3in}
\begin{tabular}{ || c | c | c | c | c  ||}
\hline \hline 
$\mathsf{g}$ & $\mathsf{k}$ & $m$ & $d$ & Conditions  \\ \hline  \hline 
$A_r$ & $A_{r-2s} \oplus^{{\!\!\!\!}^{t}} u(1)$ & 
         $t$& $4s(r-s+1)$& $ r \geq 3 \ , \ 
         1 \leq s \leq [\frac{r-1}{2}] \ , \ 0 \leq t \leq s $ \\  \hline 
$ A_{2r}  $ & $  \oplus^{{\!\!\!\!}^{s}} \, u(1)$
          & $ s $ & $4r(r+1) $ & $ r \geq 1 \ , \ 
         0 \leq s \leq r  $ \\  \hline
$ A_{2r-1}  $ & $ \oplus^{{\!\!\!\!}^{s}} \, u(1)  $ 
         & $ s+1 $ & $4r^2 $ & $r \geq 1 \ , \  
         0 \leq s \leq r-1$ \\  \hline
$ B_r  $ & $ B_{r-2s} \oplus^{{\!\!\!\!}^{t}} A_1 $ & $2s -t $ 
         & $4(s(2r -2s +1)-t)$
         & $r \geq 3\ , \ 1 \leq s \leq [\frac{r-1}{2}] \ , \ 0 \leq t
         \leq s $ \\ \hline
$ B_r  $ & $ \{ 0\}$ & $r $ & $2r(r+1) $ & $r \geq \ 3, \  $ \\  \hline
$ B_{2r}  $ & $  \oplus^{{\!\!\!\!}^{s}} \, A_1   $ 
         & $2r - s  $ & $4(r(2r+1)-s) $ & $r \geq 2 
         \ , \  0 \leq s \leq r$ \\  \hline
$ C_r  $ & $ C_{r-s}$ & $s $ & $2s (2r -s+1) $ & $r \geq 2\ , \  
         1 \leq s \leq r-1$ \\ \hline
$ C_r  $ & $ \{0\} $ & $ r $ & $2r(r+1) $ & 
         $r \geq \ 2  $ \\  \hline
$ D_r  $ & $ D_{r-2s} \oplus^{{\!\!\!\!}^{t}}  A_1 $ 
         & $ 2s -t $ & $4(2s(r-s)-t)$
         & $r \geq 5\ , \ 1 \leq s \leq [\frac{r-3}{2}] \ , \ 0 \leq t \leq
         s $ \\  \hline
$ D_{2r}  $ & $  \oplus^{{\!\!\!\!}^{s}} \, A_1$ 
         & $ 2r -s$ & $4(2r^2-s) $ & $r \geq 2\ , \  
         0 \leq s \leq r+1$ \\  \hline
$ D_{2r +1}  $ & $ \oplus^{{\!\!\!\!}^{s}} A_1 
         \oplus^{{\!\!\!\!}^{t}}  u(1) $ & $ 2r+t-s-1
         $ & $4(2r(r+1)-s) $ & $r \geq 2\ , \ 0 \leq s \leq r 
         \ , \ 0 \leq t \leq 1$ \\  \hline
$ E_6  $ & $A_{2s+1} \oplus^{{\!\!\!\!}^{t}}  u(1) $ 
         & $t+1 $ & $ 4(19-s(s+2))$ & 
         $0 \leq s \leq 2 \ , \ 0 \leq t \leq 2-s $ \\  \hline
$ E_6  $ & $\oplus^{{\!\!\!\!}^{s}} \,   u(1) $ 
         & $s+2 $ & $80 $ & $0 \leq s \leq 2 $ \\  \hline
$ E_7  $ & $D_6 $ & $1 $ & $68 $ &  \\  \hline
$ E_7  $ & $D_4 \oplus^{{\!\!\!\!}^{s}} \,  A_1$ & $2-s $ & $4(27-s)$ 
         & $0 \leq 1 \leq s$  \\ \hline
$ E_7  $ & $\oplus^{{\!\!\!\!}^{s}} \, A_1 $ 
         & $7-s $ & $4(35-s) $ & $0 \leq s \leq 4 $ \\  \hline
$ E_8  $ & $ E_7 $ & $1 $ & $ 116$ &  \\  \hline
$ E_8  $ & $ D_6$ & $2 $ & $184 $ &  \\  \hline
$ E_8  $ & $D_4 \oplus^{{\!\!\!\!}^{s}} \, A_1 $ & $3-s $ & $ 4(56-s) $ 
         &  $0 \leq 1 \leq s$  \\  \hline
$ E_8  $ & $\oplus^{{\!\!\!\!}^{s}} \,A_1 $ 
         & $  8 - s$ & $ 4(64-s) $ & 
         $0 \leq s \leq 4  $ \\  \hline
$ F_4 $ & $ C_s $ & $4-s $ & $2(28 -s(s+1)) $ & $ 
                     1 \leq s \leq 3$ \\  \hline
$ F_4  $ & $ \{0\} $ & $ 4 $ & $56 $ & $ $ \\  \hline
$ G_2  $ & $ \oplus^{{\!\!\!\!}^{s}} \,  A_1  $ & $2-s $ 
         & $ 4(4-s)$ &  $0 \leq 1 \leq s$  \\  \hline\hline
\end{tabular}
\vspace{4mm}
\begin{center}
Table 2: Homogeneous hyper-complex and HKT spaces
$(G/K) \times^{{\!\!\!\!\!}^{m}}U(1)$ of dimension $d$ for simple groups 
$G$.
\end{center}

\begin{theorem}
All homogeneous HKT spaces constructed in Theorem \ref{theorem3}
are of the form 
\bq
\{M^0 \times M^1 \times \ldots \times M^r\}/T \ , 
\label{ha}
\eq
 where  $M^i=\{G^i/K^i\}\times^{{\!\!\!\!\!}^{m_i}}
U(1)$ for $1\leq i \leq r$ are homogeneous 
HKT spaces listed  in Table 2 for  simply-connected, simple compact groups
$G^i$, $M^0$ is a toral group and $T$ is a toral subgroup of
$B=M^0 \times^{{\!\!\!\!\!}^{m}}U(1)$ for $m=\sum_{i=1}^r m_i$.
\label{theorem4}
\end{theorem}


\section{Homogeneous QKT spaces}

The goal of this section is to construct homogeneous QKT spaces.
For this we shall use techniques   similar to those for
constructing homogeneous quaternionic spaces in \cite{joy}.
We shall show that  these homogeneous quaternionic spaces
admit QKT structures.

We begin with a homogeneous HKT manifold, $G/K$, where $G$ is a group with
reductive Lie algebra, 
$\g{g}=\g{\tilde{u}}\oplus^{{\!\!\!\!\!}^{m}} \g{u(1)}$. Our 
QKT spaces are of the form
$G/(K\times\Phi(U(2)))$, where  $\Phi$ 
is an embedding of the group $U(2)$ in $G$. This embedding is chosen
such that

\begin{description} 
\item[(i)] $\Phi(U(2))$
centralizes $K$ in $G$, 
\item[(ii)]  $\Phi(U(2))$ is a hyper-complex
sub-manifold of
 $G/K$ and 
\item[(iii)]  the left  action of $\Phi(U(2))$  on $G/K$  
induces an $SO(3)$ rotation on the three complex structures.
\end{description} 

Let us take the embedding $\Phi$ at the Lie algebra level as
\bq
\Phi(\g{u(2)}):= \mathrm{span}\{U,\phi(Y_1), \phi(Y_2),\phi(Y_3)\} 
\label{defu2}
\eq
 for the basis vectors $U:= \sum_{k=1}^{l} U_k$ and $\phi(Y_r):=
\sum_{k=1}^{l} \phi_k(Y_r)$, where the rest of the notation can be found in
section four.   The centre of $U(2)$ under the embedding $\Phi$  must be a 
\emph{closed}  subgroup of the maximal torus of $G$. For $l>1$, this places a
rationality condition on $U$ which can be satisfied for  only a dense subset of
the hyper-complex structures on $G/K$.  
The homogeneous spaces $G/(K
\times\Phi(U(2)))$ admit a quaternionic structure provided that the
hyper-complex structure on $G/K$ has been chosen such that $U$ is rational.
As we have seen  the hyper-complex structures on $G/K$ are
para\-meterized by $l^2$
\emph{real} free parameters whereas those that lead to   quaternionic 
structures on $G/(K \times\Phi(U(2)))$ are parameterized by  
$l^2$ \emph{rational} free parameters. Different
hyper-complex structures on  $G/K$  give
rise to different embeddings of $U$ in $G$, which may
lead to  quaternionic
spaces  with distinct  topological structures. We summarize the above as
\cite{joy}:

\begin{theorem}
A compact homogeneous space of the form $G/(K \times\Phi(U(2)))$ 
admits quaternionic structures if the homogeneous space $G/K$ is 
hyper-complex  and 
 $\Phi$ is an appropriate embedding of $U(2)$ in $G$.
\label{theorem6}
\end{theorem}

Let $\g{g}=\g{m}\oplus \g{k}$ be the orthogonal
decomposition of $\g{g}$ with respect to the invariant metric on
$\g{g}$ associated to the homogeneous HKT space $G/K$.
Then the  quaternionic 
space  of the form $G/(K\times\Phi(U(2)))$ admits  QKT structures
provided that 

\begin{description}

\item [(i)] the invariant metric on $\g{g}$
decomposes orthogonally as $\g{g}=\tilde{\g{m}}\oplus \tilde{\g{k}}$, 
where $\tilde{\g{k}}=\g{k}\oplus\Phi(\g{u(2)})$, 

\item [(ii)]
the invariant metric on $\tilde{\g{m}}$ is tri-hermitian with respect to the
almost quaternionic structures $J_r$, where $J_r$ are  endomorphisms
of $\tilde{\g{m}}$
associated with the quaternionic structure on $G/\{K\times\Phi(U(2))\}$,

\item [(iii)] the
torsion (\ref{torsion}) is (2,1) and (1,2) with respect to  $J_r$ and 

\item [(iv)] the
conditions for the  metric on $\g{m}$ 
to be tri-hermitian are
 compatible with the rationality condition for
the embedding of the centre of $\g{u(2)}$ in $\g{g}$.

\end{description}
 
It is  sufficient to discuss the  decomposition of 
the invariant metric of $\g{g}$  on  the subspace 
$\oplus^{{\!\!\!\!}^{l}}_{{\!\!\!\!\!\!\!\!}_{_{k=1}}} \!  (\g{d}_k
\oplus \g{u(1)}_k)$,
since $\Phi(u(2))\subset 
\oplus^{{\!\!\!\!}^{l}}_{{\!\!\!\!\!\!\!\!}_{_{k=1}}} \!  (\g{d}_k
\oplus  \g{u(1)}_k)$. We take $\g{u(2)}_k=\mathrm{span} \{U_k, \phi_k(Y_1),
\phi_k(Y_2), \phi_k(Y_3)\}$, and relabel $U_k=T^k_0$ and $\phi_k(Y_r)=T^k_r$,
$r=1,2,3$.  Let $\Phi(\g{u(2)})=\mathrm{span} \{K_a; a=0,1,2,3\}$, then 
\bq
K_a=T_a^1+T_a^2+\cdots+T_a^l \, ,  \ \ 
M_a^n=\sum_{q=1}^n \frac{T_a^q}{B(T_a^q,T_a^q)} -
\frac{n\, T_a^{n+1}}{B(T_a^{n+1},T_a^{n+1})} \  
\label{mai}
\eq
 is an orthogonal basis in 
$\oplus^{{\!\!\!\!}^{l}}_{{\!\!\!\!\!\!\!\!}_{_{k=1}}} 
\! \g{u(2)}_k$, where $1\leq n\leq
l-1$. This can be used to induce an invariant metric in
the complement of $\Phi(u(2))$  as required.

As we have seen, $\g{m}=\Phi(\g{u(2)})\oplus \tilde\g{m}$. 
From  $I_r (\Phi(\g{u(2)}))\subset \g{u(2)}$ it is easy to deduce 
that $I_r (\tilde \g{m})\subset \tilde \g{m}$. Then the 
endomorphisms $J_r$ of $\tilde \g{m}$ associated with the 
quaternionic structure on $G/(K\times \Phi(U(2))$
are defined as $J_r=I_r|_{\tilde \g{m}}$. Hence the hermiticity of 
$B$ with respect to $J_r$ is implied by the hermiticity of $B$ with
respect to $I_r$.

The three-form on $\tilde \g{m}$ associated with the torsion of the canonical
connection of  $G/(K\times
\Phi(U(2))$ is the restriction of that on $\g{m}$ associated with the HKT
structure on $G/K$. Using that the torsion of the HKT structure is (1,2) and
(2,1) with respect to $I_r$ and $I_r (\tilde
\g{m})\subset \tilde \g{m}$, we deduce that the torsion of $G/(K\times
\Phi(U(2))$ is also (1,2) and (2,1) with respect to $J_r$. 

It remains to investigate the rationality condition for $l>1$ of the
embedding of the centre of $U(2)$ in $G$. There are two cases
to consider depending on whether or not $U$ involves  a linear
combination of Cartan sub-algebra generators 
of the semi-simple sub-algebra $\g{\tilde{u}}$
of $\g{g}$. If $U$ does not involve  a linear combination of
Cartan sub-algebra generators, it is always possible to find a rational
basis such that $U$ generates a closed subgroup of $G$ and  the hermiticity
conditions of the
invariant metric on $\g{m}$ are satisfied. If $U$ involves linear combinations
of Cartan sub-algebra generators of $\g{\tilde{u}}$, then they  
always arise from the
decomposition of an $A_r$ sub-algebra of $\g{\tilde{u}}$. 
In this case, we can again
choose a basis such that $U$ has the desired properties. To see this,
we begin with the basis of the relevant Cartan sub-algebra generators found in
(\ref{restc1}) and (\ref{restc3b}), which is irrational. However,
we can use an orthogonal transformation which preserves the inner product
\bq
B(U_j,U_k)= c \, \delta_{j,k} \ 
\label{diag}
\eq
 to find another basis $\{ \tilde U_k\}$, which is rational,  such that 
$\tilde U=\sum _k \tilde U_k$
 generates a closed subgroup in $G$. To see this, we observe that 
the orthogonal
transformation has $\frac{l (l-1)}{2}$ free parameters and the rationality
of $\tilde U$ requires at most $l-1$ conditions. Hence a solution can always
be found.  

The endomorphisms $J_r$ are sections of a rank three associated bundle $V$ of a
principal
$\Phi(Sp(1))/\bZ_2$-bundle over $G/(K\times
\Phi(U(2))$. Moreover they are covariantly constant with respect to the
canonical connection $\omega$ on $G/(K\times
\Phi(U(2))$. The connection of the bundle $V$ is given by the $Sp(1)$
component of $\omega$, i.e. the connection $B_r$
in (\ref{aoneb}) is the $\Phi(\g{so(3))})$ component of the 
canonical connection:

\begin{theorem}
All homogeneous  spaces of the form 
$(M^0\times M^1 \times \ldots \times M^r)/(T\times \Phi(U(2))$
admit QKT structures if the homogeneous spaces 
$(M^0\times M^1 \times \ldots \times M^r)/T$ admit HKT structures. 
\label{theorem5}
\end{theorem}

We remark  that the topological product 
$M_1\times M_2$ of two QKT manifolds $N^1$ and $N^2$ 
does not admit a QKT structure.  Nevertheless there is 
a notion of a topological 
product of two  QKT manifolds $N^1$ and $N^2$, which we call 
\emph{join} and denote by
$N^1*N^2$. This product is similar to  that of two quaternionic manifolds
given in \cite{joy}. This operation has been used implicitly in the
poof of the  above theorem.

A simple example of a manifold with a QKT structure is $U(2)$. For this,
we observe that $U(2)$ is an HKT manifold. Then we take the product
$U(2)\times U(2)$ and embed $U(2)$ diagonally in $U(2)\times U(2)$.
Applying the above theorem, we conclude that 
$M=(U(2)\times U(2))/U(2)$ admits a QKT structure.
In fact the torsion vanishes and $M$ is a QK manifold.
However, $M$ is diffeomorphic to $U(2)$, so $U(2)$ admits both an HKT and
a QK structure. Along the same lines 
 it is possible to define a QKT structure on
the homogeneous space $(\times^{{\!\!\!\!}^{l}} \, U(2))/U(2)$, which is 
diffeomorphic to 
$\times^{{\!\!\!\!\!\!\!}^{l-1}} U(2)$ and which also admits an HKT structure.

In Table 3 we   list  all eight-dimensional homogeneous 
HKT spaces and their
associated four-dimensional homogeneous QKT spaces (up to possibly finite
coverings). We remark that  the embedding of $U(1)$ 
in $SU(3)\times U(1)$
is  parametrized by a rational number, which gives rise to  HKT
spaces with different topology. However, they  lead to the same QK
space $\bC P^2$.

\vspace{6mm}
\hspace{0.1in}
\begin{tabular}{|| c | c | c  ||}
\hline \hline 
$G/K$ & $G/\{K\times \Phi(U(2))\}$ & Comment  \\ \hline  \hline 
$SU(3)$ & $\bC P^2$ & Wolf space \\ \hline 
$\{SU(3)\times U(1)\}/U(1)$ & $\bC P^2$ & Wolf space \\ \hline 
$\{Sp(2)/Sp(1)\}\times U(1)$& $S^4$&  Wolf space \\ \hline 
$U(2)\times U(2)$ & $S^1 \times S^3$ & new QK space \\ \hline 
$U(2)\times^{{\!\!\!\!}^{4}}\ U(1)$ & - & - \\ \hline 
$\times^{{\!\!\!\!}^{8}}\ U(1)$& $\times^{{\!\!\!\!}^{4}}\ U(1)$& flat
space  \\ \hline
\end{tabular}
\vspace{2mm}
\begin{center}
Table 3: Homogeneous HKT spaces $G/K$ of dimension eight and  associated 
homogeneous QK  spaces $G/(K\times \Phi(U(2))$. 
\end{center}
\vspace{4mm}

Our construction of homogeneous QKT spaces includes that of QK spa\-ces.
In fact, only the first level of the above decomposition of $G$, i.e. $l=1$
is required for the construction of homogeneous  QK spaces. In particular, the
Wolf spaces are found in this way from the decomposition of simple Lie
algebras. Note that compact Wolf spaces have 
non-compact duals, as there are no Cartan sub-algebra generators
left in $\tilde{\g{m}}$, but this is not so
for generic homogeneous QKT spaces.


\section{Twistor spaces for homogeneous QKT spaces}

The holonomy of a  QKT manifold $M$ is a subgroup of $Sp(d)\cdot Sp(1)$.
Therefore, its tangent bundle is associated to 
a principal $Sp(d)\cdot Sp(1)$-bundle.
The complexification of the tangent space of $M$ can be
split as $T^cM=T_{2d}\otimes T_2$
with the first sub-bundle associated with $Sp(d)$ 
and the second one associated with $Sp(1)$.
This structure group can be lifted to $Sp(d)\times Sp(1)$ provided
that the second Witney class of $M$ vanishes, i.e. if $M$ is a spin
manifold. The twistor space $Z$ of a QKT manifold \cite {qkt} can be 
defined as the
projectivization of $T_2$ and it has been shown 
that it is  a complex manifold. Using the conventions and notation
of \cite{qkt}, we show:

\begin{theorem}
The twistor space $Z$ for any QKT manifold of dimension 
bigger than four admits a KT 
structure.
\end{theorem}

\no To prove this theorem
it is sufficient to show that there is a non-degenerate
(1,1) form on $Z$ with respect to the complex structure in $Z$. Such a form is
\bq
\Omega=2i \left(E_1{}^2\wedge E_2{}^1+E^{b2}\wedge E^{a1} \eta_{ab}\right)\ ,
\eq
where $\{E_1{}^2, E^{a2}\}$ is a basis of 
(1,0)-forms, $\{E_2{}^1, E^{a1}\}$
is a basis of (0,1)-forms and $\eta$ is the invariant 
symplectic form of $Sp(d)$.
The metric and the torsion of the KT structure on $Z$ can be determined 
from the complex structure and $\Omega$. 

In \cite{qkt} it was also shown that if the exterior derivative 
$dH$ of the torsion $H$
on $M$ is (2,2) with respect to all endomorphisms $J_r$ and if a 
certain non-degeneracy
condition is met, then the twistor space is  a K\"ahler manifold but 
with respect
to a different metric from the one given above. For homogeneous
QKT manifolds one can show:

\begin{theorem}
Homogeneous QKT spaces with $dH$ a (2,2) form  are four-dimensional.
\label{twotwo}
\end{theorem}

\no To prove this we use 
\bq
[f_r,f_s]=2\epsilon_{rs}{}^t f_t \, , \  \
[f_r,J_s]=2\epsilon_{rs}{}^t J_t \label{f2} \,  , \ \
[J_r,J_s]=2\epsilon_{rs}{}^tJ_t \, , 
\label{fj}
\eq
where $f_r$ is the representation of
the $\g{sp(1)}$-part of $\g{k}$
on $\g{m}$ (\ref{fj}). We choose a frame 
compactible with respect
to one of the almost complex structures, say  $J_1$, i.e.    
$(J_1)_{\mu}{}^{\nu}=i\delta_{\mu}{}^{\nu}$. Using the above
relations, we find
$(J_2)_{\mu}{}^{\nu}=(J_3)_{\mu}{}^{\nu}=0$.
Putting these into the second equation in (\ref{fj}), we find that
$(f_1)_{\mu}{}^{\bar{\nu}}=0$, $(f_2)_{\mu}{}^{\bar{\nu}}= i
(J_3)_{\mu}{}^{\bar{\nu}}$ and $(f_3)_{\mu}{}^{\bar{\nu}}= -i
(J_2)_{\mu}{}^{\bar{\nu}}$ and from the third equation in 
(\ref{fj}) we deduce that  
$(I_2)_{\mu}{}^{\bar{\nu}}= - i (J_3)_{\mu}{}^{\bar{\nu}}$. 
The (4,0)-part of $\mathrm{d}H$ is automatically zero as it
can be seen by  direct computation, whereas the (3,1)-part is 
\bq
(\mathrm{d}H)^{(3,1)}_{[\mu\nu\rho\bar{\sigma}]}=
(f_r)_{[\mu\nu}(f_r)_{\rho\bar{\sigma}]}=
(f_2)_{[\mu\nu}(J_2)_{\rho\bar{\sigma}]} 
+i(f_3)_{[\mu\nu}(J_2)_{\rho\bar{\sigma}]}\ .
\eq
For this term to vanish, 
$(f_2)_{\mu\nu}+i(f_3)_{\mu\nu}=0$ or $d=1$. Assuming  the former we 
 deduce from the
third relation in  (\ref{fj}) that $(f_1)_{\mu}{}^{\nu}=i
\delta_{\mu}{}^{\nu}$. So the structure 
constant $f_1$ are proportional
to the complex structure $J_1$. In a similar way, we can show 
that  $f_r \sim J_r$. 
This implies that the $Sp(1)$ part of the curvature
is proportional to the endomorphisms of the QKT manifold.  But
it has been shown in \cite{qkt} that such QKT manifolds have vanishing
torsion. Thus
the only QKT spaces with  non-zero torsion, 
whose exterior derivative is (2,2)
with respect to all $J_r$'s, are four dimensional.

The twistor space $Z$ of the homogeneous QKT space 
$M=G/(\Phi(U(2))\times K)$ is of
the form $Z=G/(\Phi(U(1)\times U(1))\times K)$, where
$U(1)\times U(1)\subset U(2)$. The complex structure on the
twistor space is induced by the complex structure 
in the HKT space $G/K$ 
which is invariant under the right action of 
$\Phi(U(1)\times U(1))$; the orbit of this action on the space of complex
structures is one-dimensional.

We have seen that our homogeneous QKT manifolds $M$ admit by construction
the fibration $\Phi(U(2))\rightarrow
G/K\rightarrow M$, where $G/K$ is an HKT manifold. It turns out that
every QKT manifold admit such a fibration. This 
fibration can be constructed along the same lines as the one over
quaternionic manifolds.
In order to do this,  we remove
the zero section of $T_2$ and compactify each fibre to a Hopf surface.
To find a fibration of  QKT manifolds which  
reduces to that of the homogeneous
ones above, we further have to  twist with a $U(1)$ bundle. It seems likely
that the resulting spaces admit an HKT structure.


\section{Conclusions}

We have investigated a class of KT, HKT and QKT structures on 
homogeneous spaces $G/K$  using an invariant metric on $G$
and the canonical connection. Our construction was based on an
orthogonal decomposition of $G$ which can be most easily understood
using Dynkin diagrams. Lists of KT, HKT and QKT spaces were compiled.
We have also studied the twistor spaces of homogeneous QKT  spaces
and have found that they admit a KT structure.

As we have mentioned, these geometries have appeared in the context
of sigma models and string theory.  Group spaces that admit an 
HKT structure  are
vacua of string theory. One  reason for this is that  these HKT 
geometries have torsion which
is a  harmonic three-form with respect to the invariant metric. 
This is no longer
the case for our homogeneous HKT manifolds. The exterior 
derivative of the torsion
of these spaces can be written as the trace of the square of 
the curvature of the
canonical connection. This is reminiscent of the condition for 
the cancellation
of the gravitational anomaly of the heterotic string at one 
loop in the sigma model
perturbation theory. However, since there is no `classical' torsion which
is a closed three form associated with this geometry. The only 
way to make sense
of this is to assume that the one loop anomaly cancellation condition is
exact and that the string tension has a particular value for 
this background.
It may be interesting to investigate this further in the future.
In connection with M-theory, it is worth pointing out that our HKT
eight-dimensional manifolds are closely associated with 
some of the Rubin-Freud spaces.  In particular, most eight-dimensional
homogeneous  HKT spaces 
are of the form $M_{(8)}=M_{(7)}\times U(1)$ (see Table 3),
where $M_{(7)}$ are Freud-Rubin spaces \cite{freund, cast}, which are special
seven-dimensional Einstein spaces. 

It would be of interest to develop techniques to
construct systematically non-homogeneous KT, HKT and
QKT spaces. Some examples of KT and HKT spaces are known but 
all of them are non-compact.
In particular, the QKT spaces which satisfy
the requirements of the theorem in \cite{qkt} will lead
to the construction of HK manifolds. 

\vskip 1cm
\noindent{\bf Acknowledgments:} We thank P. S. Howe and
D. Joyce for helpful discussions. 
 A.O. is supported by the EPSRC and the German
 National Foundation. G.P. is supported by a
University Research Fellowship from the Royal Society.
\vskip 1cm



\end{document}